\documentclass{appolb}
\usepackage{epsfig}
\usepackage{amsmath}
\usepackage{graphicx}

\begin{document}

\title{Geometrical scaling in hadronic collisions\thanks{%
Based on a talk presented at Cracow Epiphany Conference "On the First Year
of the LHC", Krakow, January 10-12, 2011.}}
\author{Michal Praszalowicz\thanks{e-mail: {\tt michal@if.uj.edu.pl}}
\address{
M. Smoluchowski Institute of Physics,  \and  Jagellonian University,
 \and  Reymonta 4, 30-059 Krakow, Poland} }
\maketitle

\begin{abstract}
We show that the $p_{\mathrm{T}}$ spectra measured in pp collisions  at the
LHC exhibit geometrical scaling introduced earlier in  the context of deep
inelastic scattering. We also argue that  the onset of geometrical scaling
can be seen in nucleus-nucleus  collisions at lower RHIC energies.
\end{abstract}


\section{Introduction}

\label{intro}

With the start of the LHC we have been confronted with a wealth of data on
multiparticle production at high energies both in pp \cite{Aamodt:2009dt}--%
\cite{Aad:2010rd} and in heavy ion collisions \cite{Aamodt:2010pb}. One of
the remarkable results is that total multiplicity of charged particles
produced in central rapidity in pp collisions is rising like a power of $s$%
\begin{equation}
\frac{dN_{\text{ch}}}{d\eta}\sim s^{\tilde{\lambda}} \;\text{with}\;\tilde{%
\lambda}\simeq 0.23.   \label{srise}
\end{equation}
The power law behavior (\ref{srise}) is expected in saturation
models \cite{Gribov:1984tu}--\cite{Ayala:1995hx}. In this paper we show that
power like behavior of total multiplicity follows naturally if $p_{\text{T}}$
spectra of charged particles exhibit geometrical scaling \cite%
{McLerran:2010ex,Praszalowicz:2011tc}. In Sect.~\ref{DIS} we remind basic
properties of geometrical scaling which was introduced in the context of
small $x$ deep inelastic scattering (DIS). In Sect.~\ref{sectpp} we
introduce geometrical scaling in pp collisions. In Sect.~\ref{sectHI} we
briefly discuss a possibility of geometrical scaling in heavy ion
collisions. Finally we summarize and give conclusions in Sect.~\ref{concl}.

\section{Geometrical scaling in DIS}

\label{DIS}

In a successful description of small $x$ DIS proposed in seminal papers by
Golec-Biernat and W\"{u}sthoff (GBW model) \cite{GolecBiernat:1998js}, a
cross-section for virtual photon-proton scattering in DIS reads:%
\begin{equation}
\sigma_{\gamma^{\ast}p}=\int dr^{2}\left\vert \psi(r,Q^{2})\right\vert
^{2}\sigma_{dP}(r^{2}Q_{\text{s}}^{2}(x)).   \label{redsig}
\end{equation}
Here $\psi$ is the wave function describing dissociation of a virtual photon
into a $q\bar{q}$ dipole and $\sigma_{dP}$ is a dipole-proton cross-section.
The main assumption of the GBW model is that $\sigma_{dP}$ which in
principle is a function of two independent variables: dipole size $r $ and
dipole-proton energy $W$ (or Bjorken $x$), depends in practice only on a
certain combination of these two variables, namely on the product $r^{2}Q_{%
\text{s}}^{2}(x)$ where%
\begin{equation}
Q_{\text{s}}^{2}(x)=Q_{0}^{2}\left( \frac{x_{0}}{x}\right) ^{\lambda}
\label{Qsat}
\end{equation}
is called a saturation scale. Bjorken $x$ is defined as%
\begin{equation}
x=\frac{Q^{2}}{Q^{2}+W^{2}}.
\end{equation}
Here $Q_{0}\sim1$ GeV and $x_{0}\sim10^{-3}$ are free parameters whose
precise values can be extracted by fitting (\ref{redsig}) to the HERA data.
Power $\lambda$ is known to be of the order $\lambda\sim0.2\div0.3$.

For transverse photons (neglecting quark masses):%
\begin{equation}
|\psi_{T}(r,Q^{2})|^{2}=\int_{0}^{1}dz\left[ z^{2}+(1-z)^{2}\right]
\overline{Q}^{2}K_{1}^{2}(\overline{Q}r)   \label{psiT}
\end{equation}
where
\begin{equation}
\overline{Q}^{2}=z(1-z)Q^{2}
\end{equation}
and $K_{1}$ is a modified Bessel function. From Eq.~(\ref{psiT}) it
follows
that%
\begin{equation}
|\psi_{T}(r,Q^{2})|^{2}=Q^{2}|\tilde{\psi}_{T}(rQ)|^{2}
\end{equation}
where we have explicitly factored out $Q^{2}$. Defining new variable $%
u=Q^{2}r^{2}$, new function $\phi(u)=u\left\vert \tilde{\psi}%
_{T}(u)\right\vert ^{2}$ and \emph{scaling variable} $\tau$
\begin{equation}
\tau=\frac{Q^{2}}{Q_{\text{s}}^{2}(x)}.
\end{equation}
we arrive at:%
\begin{equation}
\sigma_{\gamma^{\ast}p} =\int\frac{du}{u}\phi(u)\sigma_{dP}(u \tau).
\end{equation}
It follows that $\sigma_{\gamma^{\ast}p}$ is a function of scaling variable $%
\tau$, rather than a function of two variables $Q^2$ and $x$. This
phenomenon is known as geometrical scaling (GS)\cite{Stasto:2000er}. GS has
been observed in DIS data for $x<0.01$.

In the Golec-Biernat--W\"{u}sthoff model%
\begin{equation}
\sigma_{dP}(r^{2}Q_{\text{s}}^{2}(x))=\sigma_{0}\left( 1-\exp(-r^{2}Q_{\text{%
s}}^{2}(x))\right)
\end{equation}
where $\sigma_{0}$ is dimensional constant; $\sigma_{0}\simeq23$ mb.

\begin{figure}[h!]
\centering
\includegraphics[scale=0.85]{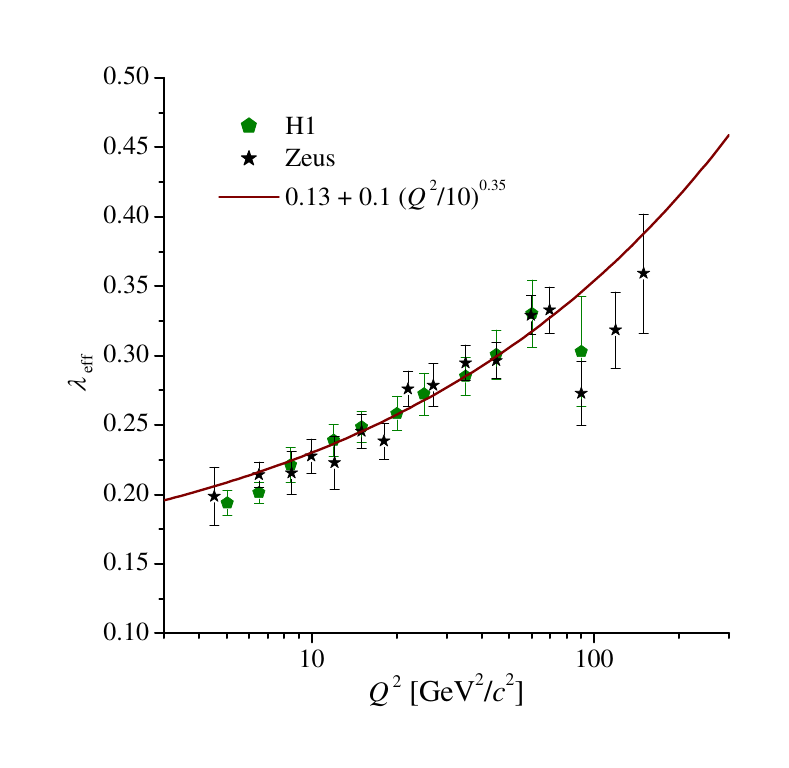}
\caption{Dependence of $\protect\lambda_{\mathrm{eff}}$ on $Q^{2}$ from HERA
(HERA data points \protect\cite{HERAdata} after Ref.\protect\cite%
{Kowalski:2010ue}). }
\label{lamHERA}
\end{figure}

In practice $Q_{\text{s}}^{2}$ may also have some residual dependence on $%
Q^{2}$ if DGLAP evolution in (\ref{redsig}) is taken into account \cite%
{Bartels:2002cj}. Indeed, it can be shown that in the GBW model -- up to
logarithmic corrections -- DIS structure function is proportional to
$Q_{\text{s}}^{2}$:

\begin{equation}
\sigma_{\gamma^{\ast}p}(x,Q^{2})\sim\sigma_{0}\frac{Q_{\text{s}}^{2}(x)}{%
Q^{2}}\;\text{and}\;F_{2}(x,Q^{2})\sim\sigma_{0}Q_{\text{s}}^{2}(x).
\label{sigF2}
\end{equation}
At first sight Eq.~(\ref{sigF2}) may look contradictory, since left hand
sides depend on $Q^{2}$ and the right hand sides do not. In practice,
exact calculation of the integral in (\ref{redsig}) renders some mild
$Q^{2}$
dependence of the right hand sides. Moreover DGLAP evolution introduces $%
Q^{2}$ dependence of $Q_{\text{s}}^{2}(x)$. Therefore the effective
saturation scale can be conveniently parameterized as%
\begin{equation}
Q_{\text{s,eff}}^{2}(x,Q^{2})=Q_{0}^{2}\left( \frac{x_{0}}{x}\right)
^{\lambda_{\text{eff}}(Q^{2})}.
\end{equation}
Exponent $\lambda_{\text{eff}}(Q^{2})$ has been extracted from the
HERA data \cite{HERAdata}. This is shown in Fig.~\ref{lamHERA} (after Ref.%
\cite{Kowalski:2010ue}) together with an eyeballing fit \cite%
{Praszalowicz:2011tc}:%
\begin{equation}
\lambda_{\text{eff}}(Q)=0.13+0.1\left( \frac{Q^{2}}{10}\right) ^{0.35}.
\label{lam}
\end{equation}

\section{Geometrical scaling in pp collisions}

\label{sectpp}

In pp collisions we do not have a viable model of low and medium $p_{\text{T}%
}$ particle production at high energies. Nevertheless one often uses a $k_{%
\text{T}}$ factorized form of a cross-section describing production of a $p_{%
\text{T}}$ gluon at rapidity $y$ \cite{Gribov:1981kg}:%
\begin{equation}
E\frac{d\sigma}{d^{3}p}=\frac{3\pi}{2}\frac{1}{p_{\text{T}}^{2}}\int dk_{%
\text{T}}^{2}\alpha_{\text{s}}(k_{\text{T}})\varphi_{1}(x_{1},k_{\text{T}%
}^{2})\varphi_{2}(x_{2},(k-p)_{\text{T}}^{2})   \label{sigG}
\end{equation}
where
\begin{equation}
x_{1,2}=\frac{p_{\text{T}}}{\sqrt{s}}e^{\pm y}   \label{x12}
\end{equation}
are Bjorken $x$'s of colliding partons. Here $\varphi$'s are unintegrated
gluon densities. Introducing "regular" gluon distribution%
\begin{equation}
xG(x,Q^{2})=\int \limits^{Q^{2}}dk_{\text{T}}^{2}\varphi(x,k_{\text{T}%
}^{2}).   \label{glue}
\end{equation}
one obtains for $p_{\text{T}}^{2}>Q_{\text{s}}^{2}$%
\begin{equation}
E\frac{d\sigma}{d^{3}p}=\frac{3\pi}{2}\frac{\alpha_{\text{s}}(Q_{\text{s}})}{%
p_{\text{T}}^{2}}\left\{ \varphi_{1}(x_{1},p_{\text{T}}^{2})x_{2}G(x_{2},p_{%
\text{T}}^{2})+\varphi_{2}(x_{2},p_{\text{T}}^{2})x_{1}G(x_{1},p_{\text{T}%
}^{2})\right\} .   \label{sigG1}
\end{equation}
There have been recently more involved model calculations of particle
multiplicity based on Eq.~(\ref{sigG}) \cite{Albacete:2010bs}--\cite%
{Tribedy:2010ab}.

Kharzeev an Levin proposed a simple Ansatz for unintegrated gluon
distribution \cite{Kharzeev:2001gp}:
\begin{equation}
\varphi(x,p_{\text{T}}^{2})=\frac{3\sigma_{0}}{\pi^{2}\alpha_{\text{s}}(Q_{%
\text{s}}^{2})}\left\{
\begin{array}{ccc}
1 & \text{for} & p_{\text{T}}^{2}<Q_{\text{s}}^{2}, \\
&  &  \\
Q_{\text{s}}^{2}/p_{\text{T}}^{2} & \text{for} & Q_{\text{s}}^{2}<p_{\text{T}%
}^{2}.%
\end{array}
\right.   \label{unglue}
\end{equation}
Hence, up to the logarithmic corrections due to the running coupling
constant, we arrive at geometrical scaling for the multiplicity distribution%
\begin{equation}
\frac{dN_{\text{ch}}}{d\eta d^{2}p_{\text{T}}}=\frac{1}{\sigma_{\text{inel}}}%
E\frac{d\sigma}{d^{3}p}=\frac{1}{Q_{0}^{2}}F(\tau)   \label{GSpp}
\end{equation}
where $Q_{0}\sim1$ GeV and $\sigma_{\text{inel}}$ is the inelastic
cross-section. Although we have used a very simple Ansatz (\ref{unglue}) for
unintegrated gluon distribution $\varphi$, it satisfies the generic property
that $xG(x,Q_{\text{s}}^{2})\sim Q_{\text{s}}^{2}$ which is enough for GS to
hold.

\begin{figure}[h]
\centering
\includegraphics[scale=0.76]{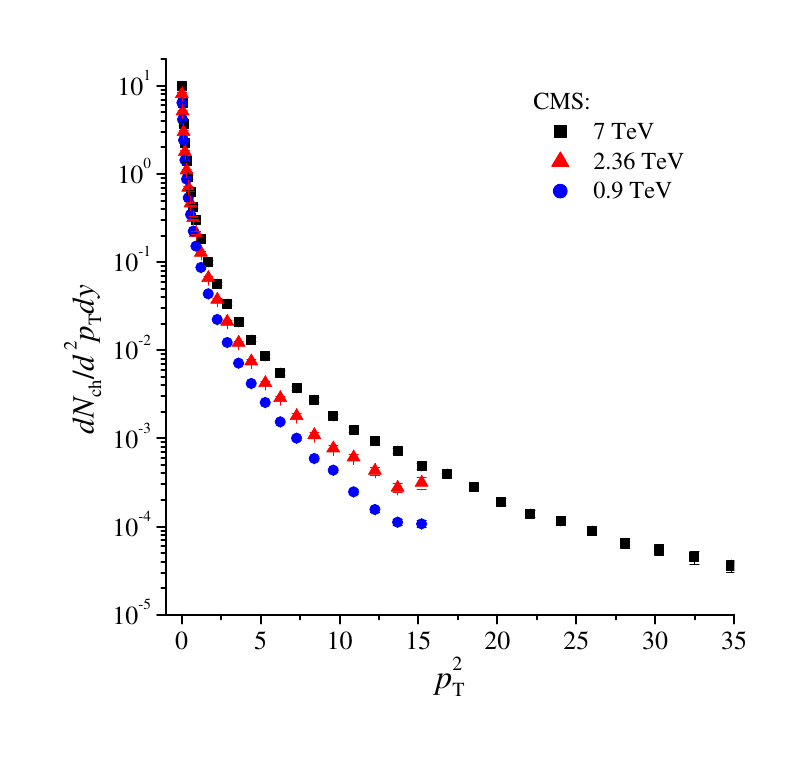} %
\includegraphics[scale=0.76]{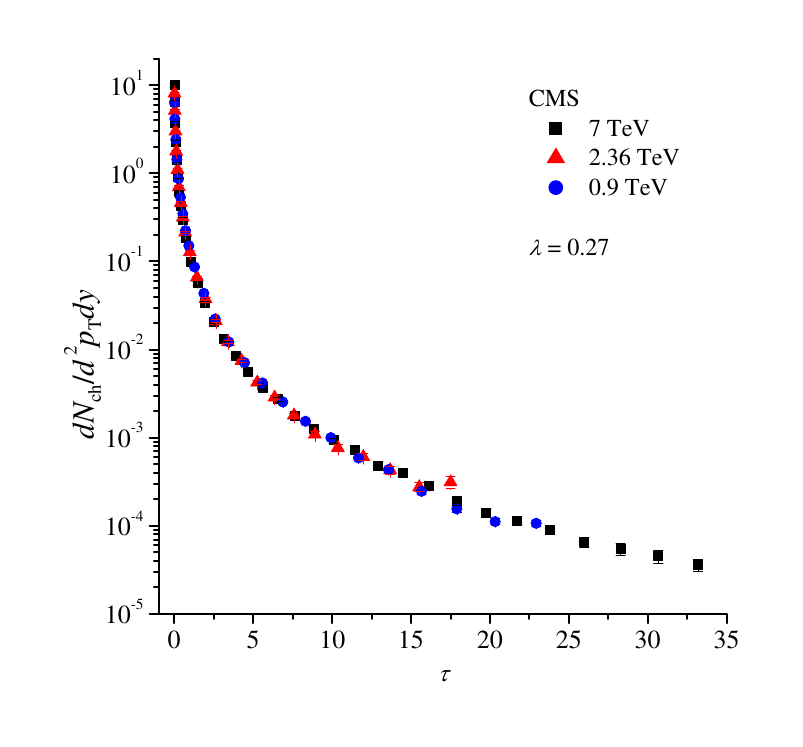}
\caption{Charged particle multiplicity at mid rapidity $|\protect\eta|<2.4$
as measured by CMS \protect\cite{Khachatryan:2010xs}, plotted as functions
of $p_{\mathrm{T}}^{2}$ (left) and scaling variable $\protect\tau$ (right)
for $\protect\lambda=0.27$. }
\label{spectraCMS}
\end{figure}

$F(\tau)$ is a universal function of the scaling variable%
\begin{equation}
\tau=\frac{p_{\text{T}}^{2}}{Q_{\text{s}}^{2}}   \label{taudef}
\end{equation}
where in view of (\ref{Qsat}) and (\ref{x12})
\begin{equation}
Q_{\text{s}}^{2}=Q_{0}^{2}\left( \frac{p_{\text{T}}}{W}\right) ^{-\lambda }
\label{Qsdef}
\end{equation}
where $W=\sqrt{s}\times10^{-3}$. Here factor $10^{-3}$ corresponds to the
(arbitrary at this moment) choice of $x_{0}$.

The power like growth of the multiplicity can be easily understood as a
consequence of geometrical
scaling. Indeed%
\begin{equation}
\frac{dN_{\text{ch}}}{dy}=\int \frac{dp_{\text{T}}^{2}}{Q_{0}^{2}}F(\tau).
\end{equation}
Simple change of variables gives \cite{McLerran:2010ex}%
\begin{equation}
\frac{dp_{\text{T}}^{2}}{Q_{0}^{2}}=\frac{2}{2+\lambda}\left( \frac{W}{Q_{0}}%
\right) ^{\frac{2\lambda}{2+\lambda}}\tau^{-\frac{\lambda}{2+\lambda}}d\tau.
\label{varchange}
\end{equation}
The integral over $d\tau$ is convergent and \textit{universal},
\textit{i.e.} it does not depend on energy. It follows from
Eq.~(\ref{varchange}) that the
effective power of the multiplicity growth is%
\begin{equation}
\tilde{\lambda}=\frac{2\lambda}{2+\lambda}<\lambda
\end{equation}
rather than $\lambda$. For $\lambda=0.27$ we have that $\tilde{\lambda}=0.238
$.

In Refs.\cite{McLerran:2010ex} it was shown that CMS charged particle $p_{%
\mathrm{T}}$ spectra \cite{Khachatryan:2010xs} at mid rapidity $|\eta|<2.4$
plotted as functions of scaling variable $\tau$ fall on one universal curve (%
\ref{GSpp}). This is depicted in Fig.~\ref{spectraCMS} where we plot $p_{%
\mathrm{T}}$ spectra for three LHC energies as functions of $p_{\mathrm{T}%
}^{2}$ (left panel) and as functions of scaling variable $\tau$ for $%
\lambda=0.27$ (right panel).

\begin{figure}[h]
\centering
\includegraphics[scale=0.77]{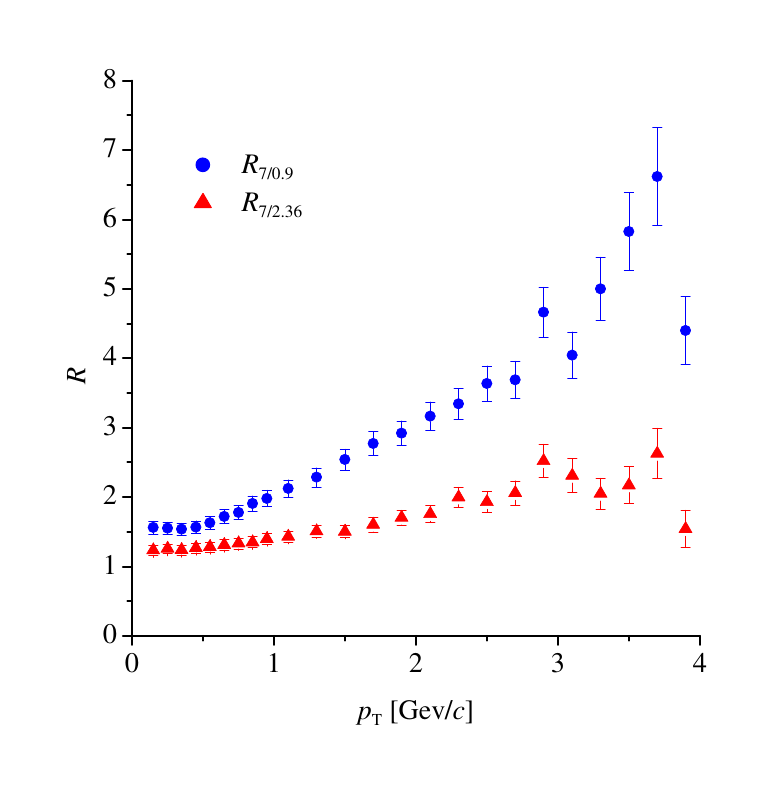} %
\includegraphics[scale=0.77]{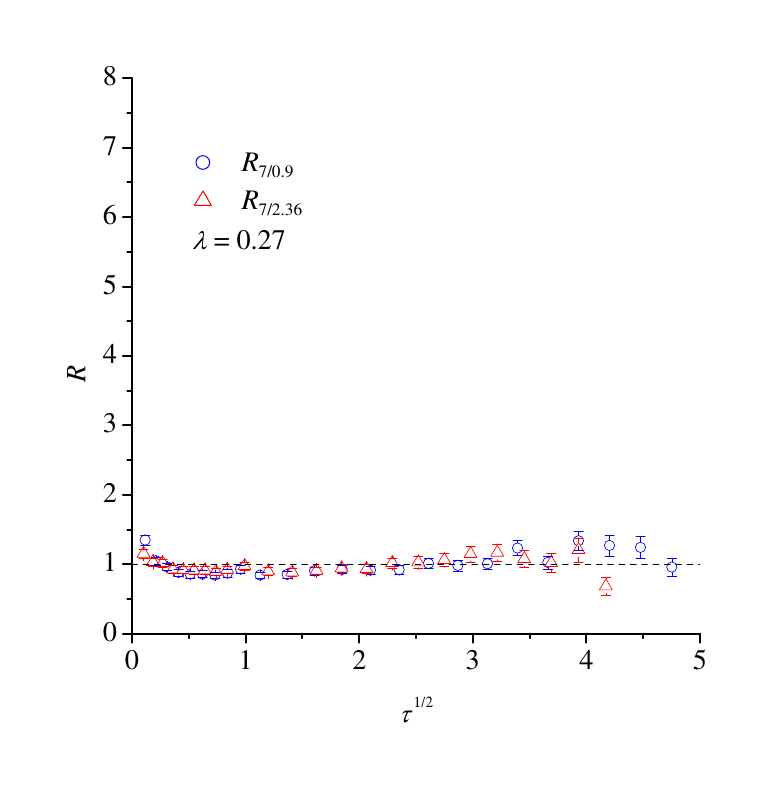}
\caption{Ratios of CMS $p_{\mathrm{T}}$ spectra \protect\cite%
{Khachatryan:2010xs} at 0.7 TeV to 0.9 (blue circles) and 2.36 TeV (red
triangles) plotted as functions of $p_{\mathrm{T}} $ (left) and scaling
variable $\protect\sqrt{\protect\tau}$ (right) for $\protect\lambda=0.27$. }
\label{ratios1}
\end{figure}

\begin{figure}[h]
\centering
\includegraphics[scale=0.77]{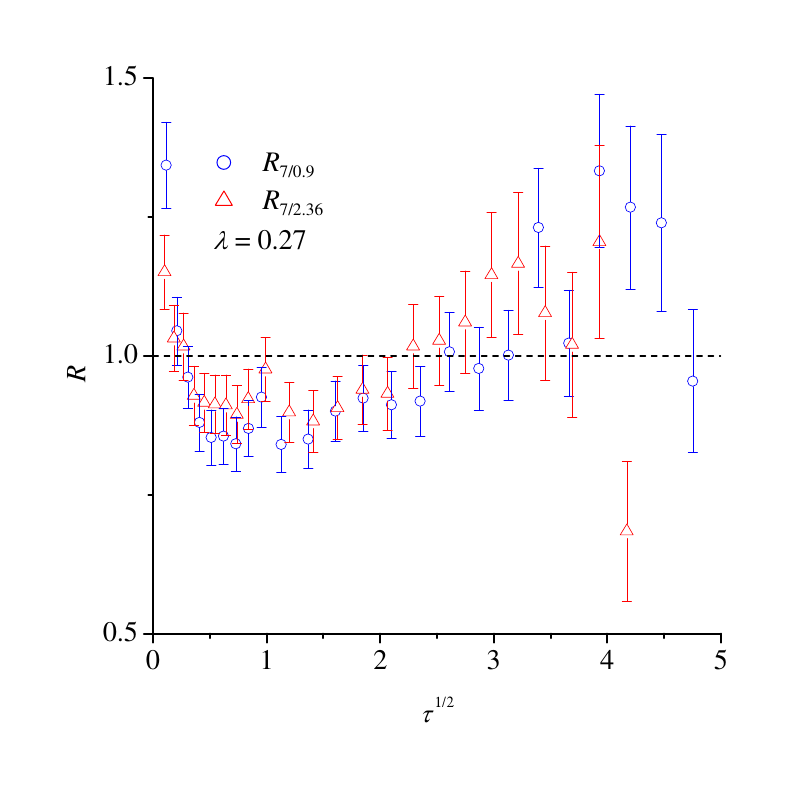} %
\includegraphics[scale=0.77]{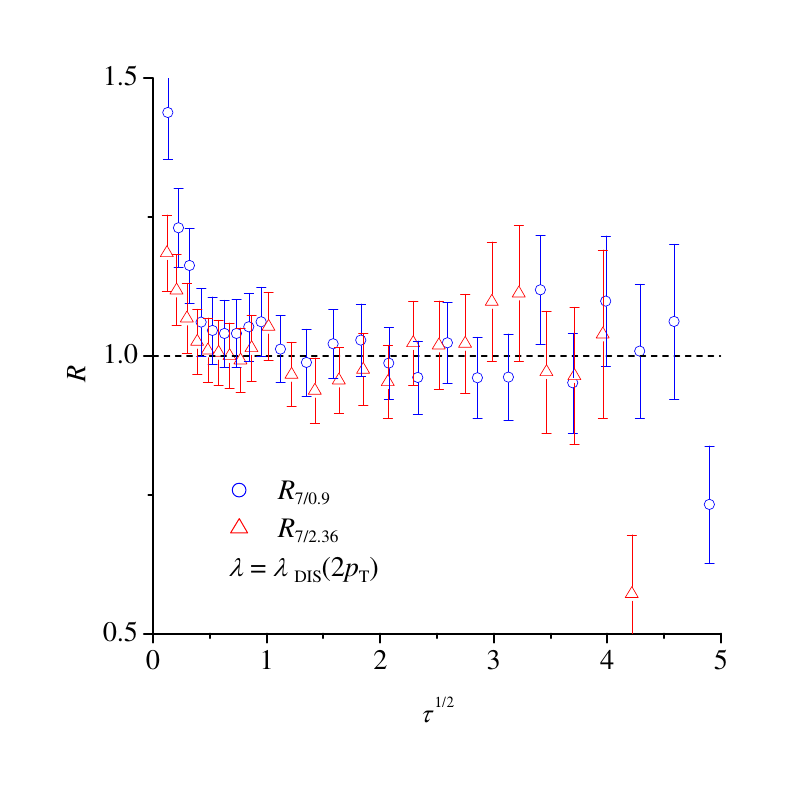}
\caption{Enlarged plot of the right panel of Fig.~\protect\ref{ratios1} for $%
\protect\lambda=0.27$ (left) and for $\protect\lambda=\protect\lambda_{%
\mathrm{eff}}(2p_{\mathrm{T}})$. }
\label{ratios2}
\end{figure}

In order to examine the quality of geometrical scaling in pp collisions we
plot in Fig.~\ref{ratios1} ratios of spectra measured at 7 TeV to spectra
at 0.9 and 2.36 TeV in function of $p_{\text{T}}$ (left panel) and
$\sqrt{\tau}
$ (right panel). We see that original ratios plotted in terms of $p_{\text{T}%
}$ range from 1.5 to 7, whereas plotted in terms of $\sqrt{\tau}$ they are
well concentrated around unity. This is further illustrated in the left
panel of Fig.~\ref{ratios2} which presents the enlarged view of the right
panel of Fig.~\ref{ratios1}. With this accuracy we see a small systematic
increase of the ratios (apart from the first 4, 5 points) which suggests
some weak dependence of exponent $\lambda$ on $p_{\text{T}}$. The value of $%
\lambda=0.27$ has been obtained by minimizing deviations of ratios $R_{7/0.9}
$ and $R_{7/2.36}$ from 1 for central $p_{\text{T}}$ points (\emph{i.e.}
rejecting first 5 and last 4 points).

Residual dependence of exponent $\lambda$ on $p_{\text{T}}$ is in agreement
with small $x$ dependence of the DIS structure function as measured in HERA
\cite{HERAdata}. In Ref.~\cite{Praszalowicz:2011tc} we have argued that this
dependence can be well approximated by use of the effective exponent $%
\lambda_{\text{eff}}$ of Eq.~(\ref{lam}) with argument $Q=2p_{\text{T}}$.
This is demonstrated in the right panel of Fig.~\ref{ratios2} where we used $%
\lambda_{\text{eff}}(2p_{\text{T}})$ to calculate the ratios $R_{7/0.9}$ and
$R_{7/2.36}$. An impressive improvement of geometrical scaling (\emph{i.e.}
of the equalities $R_{7/0.9}\simeq1$ and $R_{7/2.36}\simeq1$) can be indeed
seen.

\section{Onset of geometrical scaling in heavy ion collisions}

\label{sectHI}

Heavy ions provide much reacher information on the characteristics of
particle production at high energies. Indeed, one can study not only energy
dependence but also atomic number $A-$dependence, rapidity dependence (at
RHIC much larger rapidity range has been covered than so far at the LHC) and
finally centrality dependence. The production of quark-gluon plasma and its
ability to "remember" the initial conditions of the saturated gluonic matter
are here of primary interest. Unfortunately RHIC energies are presumably too
low for geometrical scaling to work. Nevertheless we show below, that
approximate GS can be seen in the RHIC data. To this end we choose the
PHOBOS $p_{\text{T}}$ distributions measured in gold-gold and copper-copper
collisions at 62.4 and 200 GeV per nucleon \cite{Back:2004ra,Alver:2005nb}.

\begin{figure}[h]
\centering
\includegraphics[scale=0.77]{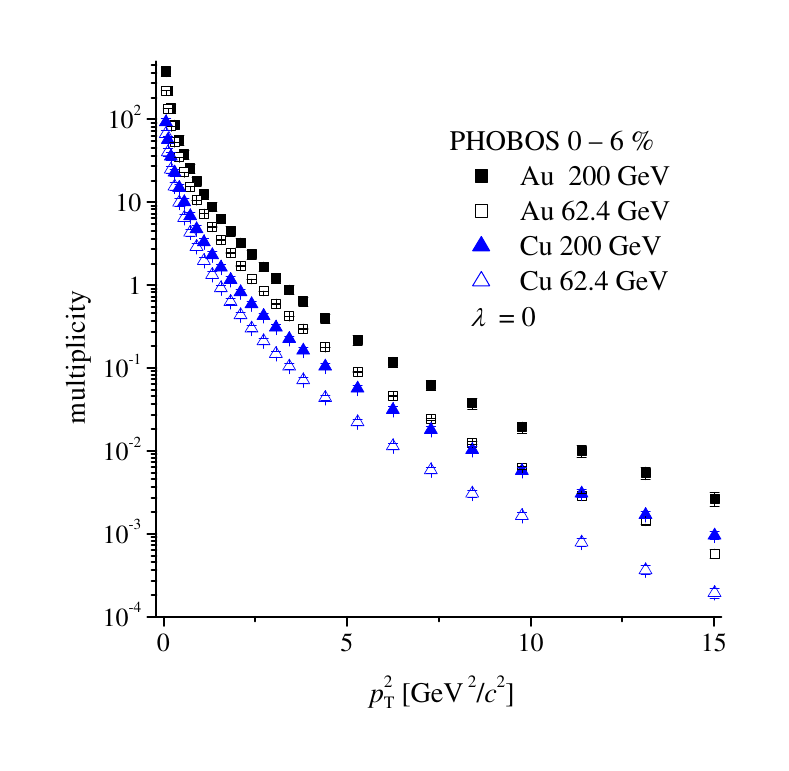} %
\includegraphics[scale=0.77]{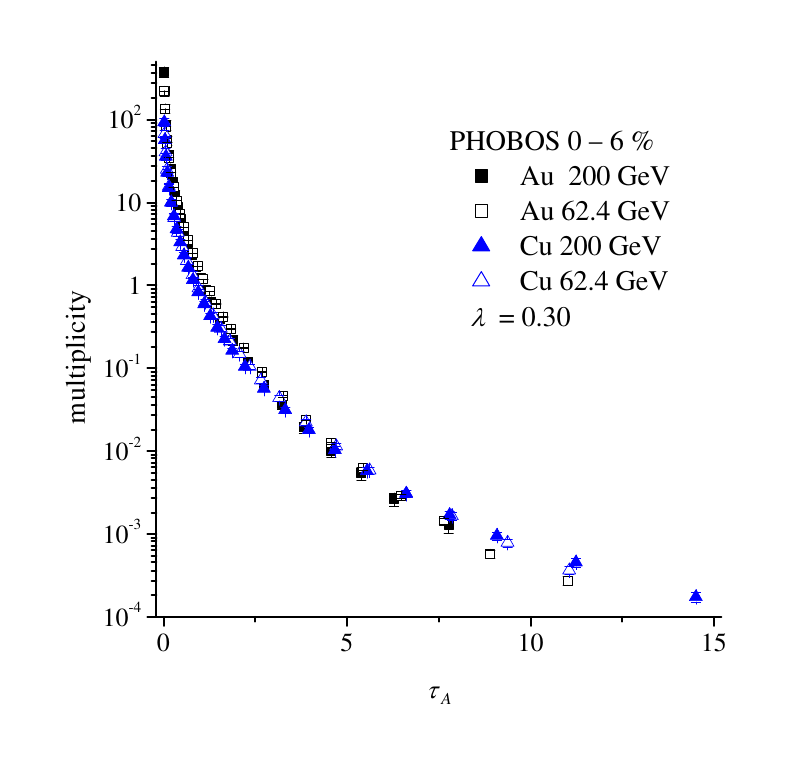}
\caption{Multiplicity distribution in heavy ion collisions for Au-Au and
Cu-Cu at two RHIC energies 200 and 62.4 GeV \protect\cite%
{Back:2004ra,Alver:2005nb} plotted in terms of $p_{\mathrm{T}}^{2}$ (left
panel) and scaling variable $\protect\tau_{A}$ (right panel). }
\label{HI}
\end{figure}

Here a new scaling law is particularly interesting. Namely the saturation
scale in nucleus-nucleus collisions scales with $A$ as \cite{Kharzeev:2000ph}
(for review see Ref.~\cite{McLerran:2010ub}):%
\begin{equation}
Q_{A\,\text{s}}^{2}=A^{1/3}Q_{\text{s}}^{2}
\end{equation}
which implies that the relevant scaling variable reads:%
\begin{equation}
\tau_{A}=\frac{p_{\text{T}}^{2}}{A^{1/3}Q_{\text{s}}^{2}}=\frac{1}{A^{1/3}}%
\frac{p_{\text{T}}^{2}}{Q_{0}^{2}}\left( \frac{p_{\text{T}}}{W}\right)
^{-\lambda}.   \label{tauA}
\end{equation}
In Fig.~\ref{HI} we plot multiplicity distribution for central Au-Au and
Cu-Cu collisions in function of $p_{\text{T}}^{2}$ and $\tau_{A}$. In this
case a slightly higher value of the exponent $\lambda$ is used, namely $%
\lambda=0.3$. We see again that the rescaled spectra seem to fall on one
curve, although the alinement of Au and Cu spectra is not perfect for small
and medium values of $\tau_{A}$. Nevertheless a tendency towards geometrical
scaling is clearly seen. Similar conclusions can be drawn for more
peripheral collisions. A detailed study of the onset of geometrical scaling
in heavy ion collisions will be presented elsewhere.

\section{Conclusions}

\label{concl}

In this paper we have demonstrated that geometrical scaling originally
postulated in deep inelastic scattering \cite{Stasto:2000er} is also
exhibited by the $p_{\text{T}}$ spectra in hadronic collisions \cite%
{McLerran:2010ex,Praszalowicz:2011tc}. To this end recent CMS data \cite%
{Khachatryan:2010xs} have been analyzed and shown to scale with scaling
variable $\tau$ defined in Eqs.(\ref{taudef},\ref{Qsdef}). A simplified
model of Gribov, Levin and Ryskin \cite{Gribov:1981kg} has been used to
motivate the appearance of GS in hadronic collisions. This model can be
a'priori used to study the shape of the universal scaling function $F(\tau)$
which deserves a separate study.

A notable difference between DIS and hadronic collisions is that in DIS we
deal with totally inclusive cross-section, whereas in pp both hadronization
and final state interactions play essential role. Nevertheless the imprint
of the saturation scale $Q_{\text{s}}$ is visible in the spectra, which
means that the information on the initial fireball survives until final
hadrons are formed.

It has been shown that the quality of geometrical scaling is improved if the
exponent $\lambda$ becomes $p_{\text{T}}$-dependent \cite%
{Praszalowicz:2011tc} in accordance with $Q$-dependence of $\lambda_{\text{%
eff}}(Q=2p_{\text{T}})$ obtained from DIS. This is a remarkable feature that
supports the picture in which medium $p_{\text{T}}$ particles are produced
from saturated gluonic matter irrespectively of the scattering states.

If so, geometrical scaling should be also present in heavy ion collisions.
The detailed studies will be certainly carried out at the LHC. Here we have
analyzed PHOBOS data \cite{Back:2004ra,Alver:2005nb} for two RHIC energies
and for two different nuclei: gold and copper, and the onset of geometrical
scaling has been clearly seen. Interestingly, we have found that the
exponent $\lambda$ that governs geometrical scaling in heavy ion collisions
is higher than the one in pp. This is in striking agreement with the fact
that multiplicity growth with energy observed by ALICE \cite{Aamodt:2010pb}
is faster in heavy ions than in pp. Question arises to what extent the
hydrodynamical evolution of the quark-gluon plasma is going to wash out
geometrical scaling that is present in the initial state. Further studies
should also concentrate on centrality and rapidity dependence of GS.

\section*{Acknowledgments}

The author wants to thank Larry McLerran for a number of stimulating
discussions that triggered this work and Andrzej Bialas for discussion and
encouragement. Special thanks are due to Barbara Wosiek for the guidance
through the wealth of heavy ion RHIC data. Part of this work has been
completed during a short visit at CERN TH Department.

\end{document}